\documentclass[aps,showpacs,twocolumn,superscriptaddress]{revtex4}
\usepackage[dvips]{graphicx}
\usepackage{amsmath}

\begin{document}

\title{Molecular wires acting as quantum heat ratchets}

\author{Fei Zhan}
\email{fei.zhan@physik.uni-augsburg.de}
\affiliation{Institut f\"ur Physik, Universit\"at Augsburg,
Universit\"atsstr.~1, D-86135 Augsburg, Germany}

\author{Nianbei Li}
\email{nianbei.li@physik.uni-augsburg.de}
\affiliation{Institut f\"ur Physik, Universit\"at Augsburg,
Universit\"atsstr.~1, D-86135 Augsburg, Germany}

\author{Sigmund Kohler}
\email{sigmund.kohler@icmm.csic.es}
\affiliation{Instituto de Ciencia de Materiales de Madrid, CSIC,
    Cantoblanco, E-28049 Madrid, Spain}

\author{Peter H\"anggi}
\email{hanggi@physik.uni-augsburg.de}
\affiliation{Institut f\"ur Physik, Universit\"at Augsburg,
Universit\"atsstr.~1, D-86135 Augsburg, Germany}
\affiliation{Department of Physics and Centre for Computational Science
    and Engineering, National University of Singapore,
    Republic of Singapore 117542}

\begin{abstract}
We explore heat transfer in molecular junctions between two leads in
the absence of a finite net thermal bias.  The application of an
unbiased, time-periodic temperature modulation of the leads entails
a dynamical breaking of reflection symmetry, such that a
directed heat current may emerge (ratchet effect).  In particular,
we consider two cases of adiabatically slow driving, namely (i)
periodic temperature modulation of only one lead and (ii)
temperature modulation of both leads with an ac driving that
contains a second harmonic, thus generating harmonic mixing.  Both
scenarios yield sizeable directed heat currents which should be
detectable with present techniques. Adding a static thermal bias,
allows one to compute the heat current-thermal load characteristics
which includes the ratchet effect of negative thermal bias with
positive-valued heat flow against the thermal bias, up to the
thermal stop-load. The ratchet heat flow in turn generates also an
electric current. An applied electric stop-voltage, yielding
effective zero electric current flow,  then mimics a solely
heat-ratchet-induced thermopower (``ratchet Seebeck
effect''), although no net thermal bias is acting.
Moreover, we find that the relative phase between the two
harmonics in scenario (ii) enables steering the net heat current
into a direction of choice.
\end{abstract}
\date{\today}

\maketitle

\section{Introduction}

In recent years, we have witnessed the development of nano-devices
based on molecular wires
\cite{Nitzan1,Reed1,hanggiCP,kohler_pr,TaoNJ1}. One of their essential
features is that the electric current through them can be controlled
effectively. One approach to such transport control is based on
conformational changes of the molecule \cite{Reed2, Kuekes1,ZhangC1}.
Another scheme relies on the dipole interaction between the molecular
wire and a tailored laser field
\cite{Lehmann1,Lehmann2,LiGQ1,Franco1,Kohler1}. A further approach
employs gate voltages acting on the wire
\cite{YangZQ1,Ghosh1,JiangF1}. The latter allows a transistor-like
control which has already been demonstrated experimentally
\cite{XiaoK1,SunY1,XuB1}. It is therefore interesting to explore as
well the related control of heat transport.

In general, heat transport through a molecular junction involves the
combined effect of electron as well as phonon transfer processes.
Control of phonon transport is much more complicated since the phonon
number is not conserved.  Nevertheless, the field of phononics, i.e.\
control and manipulation of phonons in nanomaterials, has emerged
\cite{WangL1}.  This includes functional devices, such as thermal
diodes \cite{TD1,TD2,TD3,TD4,TD5,TD6,TD7}, thermal transistors
\cite{TT1,TT2}, thermal logic gates \cite{TLG}, and thermal memories
\cite{TM} based on the presence of a static temperature
bias. The corresponding theoretical research has been accompanied by
experimental efforts on nanosystems. In particular, solid-state
thermal diodes have been realized with asymmetric nanotubes
\cite{Ex_TD1} and with semiconductor quantum dots \cite{Ex_TD2}.

Upon harvesting ideas from the field of Brownian motors
\cite{BM1,BM2,BM3,BM4,BM5} --- originally devised for particle
transport --- a classical Brownian {\it heat engine}  has been
proposed to rectify and steer heat current in nonlinear lattice
structures \cite{BHM1,BHM2}. In the absence of any static
non-equilibrium bias, a non-vanishing net heat flow can be induced
by unbiased, temporally alternating bath temperatures combined with
nonlinear interactions among neighboring lattice sites. This so
obtained directed heat current can be readily controlled to reverse
direction. If, in addition, a thermal bias across the molecule is
applied, a heat current can then typically be directed even against
an external thermal bias. This setup is therefore rather distinct
from adiabatic and nonadiabatic electron heat pumps which involve
photon assisted transmission and reflection processes in presence of
irradiating photon sources \cite{kohler_pr,rey,arrachea}.

\begin{figure}
\centerline{\includegraphics[width=8cm]{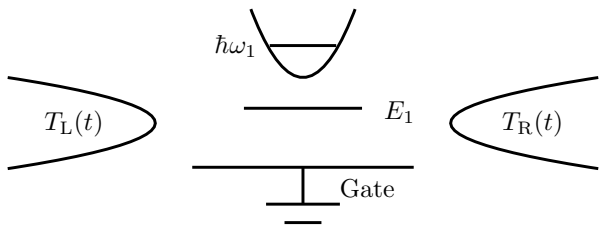}} \caption{Setup of
a molecular junction whose electronic level $E_1$ can be gated,
while the vibrational frequency $\omega_1$ is fixed. The lead
temperatures $T_{\text{L}(\text{R})}(t)$ are subjected to
time-periodic modulations.}\label{fig1}
\end{figure}

In this work, we investigate the possibility of steering heat through
a molecular junction in the presence of a gating mechanism.  In doing
so, the bath temperatures of adjacent leads are subjected to slow,
time-periodic modulations. Both the electronic and the phononic heat
current are considered, as is sketched in Fig.~\ref{fig1}. A finite
directed ratchet heat current requires breaking reflection symmetry.
This can be achieved by spatial asymmetries in combination with
non-equilibrium fields \cite{BM1,BM2,BM3,BM4,BM5} or in a purely
dynamical way \cite{SB1,SB2,SB3,SB4,SB5}. In this work we focus on
an unbiased temporal temperature variation in the
connecting leads.

This paper is organized as follows: In Section II, we specify the
physical assumptions and introduce our model together with the basic
theoretical concepts for directing heat current across a short, gated
molecular junction formed by a harmonically oscillating molecule. The
heat flux is induced by temperature modulations in the contacting
leads. Section III presents the results for case (i) where the
temperature is modulated in one lead only. We elaborate on the
phenomenon of pumping heat against a static thermal bias and consider
the resulting thermoelectric power. In Section IV, we consider case
(ii) with both lead temperatures periodically, but asymmetrically
modulated. A finite directed heat current emerges from harmonic mixing
of different frequencies, which entails dynamical symmetry breaking.
Since the leading order of this heat current is of
third order only in the driving strengths, the overall rectification
is weaker as compared to case (i) where the heat
flux starts out at second order in the driving amplitude. The latter
scheme with its symmetric static parameters, however, provides a
more efficient control scenario: The direction of the resulting heat
current can be readily reversed either by a gate voltage
or by adjusting the relative phase shift within the harmonic mixing
signal for the  temperature modulation.
Section V contains a summary and an outlook.

\section{Physical assumptions, model, and ballistic heat transfer}

We consider a molecular junction between two leads and a static gate
voltage acting on the junction.  Heat transport from both electrons and
phonons is taken into account. Since we focus on coherent
transport in a short molecular wire \cite{Nitzan,
segalnitzan2005,segal2006}, electron-phonon interaction can be
ignored. Moreover, we treat ballistic heat transfer for the electron
system and the phonon system in the absence of anharmonic interactions
and dissipative intra-wire scattering processes. Then, the heat
flux can be obtained in terms of a Landauer-type expression
involving corresponding temperature-independent transmission
probabilities for both the electrons \cite{sivan,zheng} and the
phonons \cite{ciraci,segal}. The total Hamiltonian
can thus be separated into electron and phonon part, i.e.,
\begin{equation}
 H=H^{\text{el}}+H^{\text{ph}},\label{ham}
\end{equation}
each of which consisting of a wire contribution, a lead
contribution, and an interaction term, such that
\begin{equation}
H^{\text{el}(\text{ph})}=H^{\text{el}(\text{ph})}_{\text{wire}}+H^{\text{el}(\text{ph})}_{\text{leads}}+H^{\text{el}(\text{ph})}_{\text{contacts}}\,.\label{ham-el}
\end{equation}

The short molecular wire is modeled as a single energy level and one
harmonic phonon mode only. Then the Hamiltonian of the wire electron
in tight-binding approximation reads
\begin{equation}\label{H-wire}
H^{\text{el}}_{\text{wire}}=E_1\left.|1\right>\left<1|\right.\,,
\end{equation}
where $E_1$ describes the on-site energy of the tight-binding level
which can be shifted via a gate voltage. The electrons in the
leads are modeled as ideal electron gases, i.e.,
\begin{equation}
H^{\text{el}}_{\text{leads}}=H_{\text{L}}+H_{\text{R}}=
\sum_q\epsilon_{\text{L}q}c^{\dagger}_{\text{L}q}c_{\text{L}q}+\sum_q\epsilon_{\text{R}q}c^{\dagger}_{\text{R}q}c_{\text{R}q}
\,,
\end{equation}
where $c^{\dagger}_{lq}$ creates an electron in state
$\left.|lq\right>$ of lead $l=\mathrm{\text{L},\text{R}}$. The
electron tunneling Hamiltonian
\begin{equation}
H^{\text{el}}_{\text{contacts}}=\sum_q(V_{\text{L}q}c^{\dagger}_{\text{L}q}c_1+V_{\text{R}q}c^{\dagger}_{\text{R}q}c_1)+\text{h.c.}\,,
\end{equation}
establishes the contact between the wire and the leads.
This tunneling coupling is characterized by the spectral density
$\Gamma_l(\epsilon)=2\pi\sum_q|V_{lq}|^2\delta(\epsilon-\epsilon_{lq})$.
We assume symmetric coupling within a wide-band limit such that
$\Gamma_l(\epsilon) = \Gamma$.

The phonon mode is represented by a harmonic oscillator with the
Hamiltonian
\begin{equation}
H^{\text{ph}}_{\text{wire}}=\frac{P^2}{2M}+\frac{1}{2}M\omega^2_1Q^2\,,
\end{equation}
where $Q$ and $P$ denote the position and the momentum operator,
respectively, $M$ denotes the atom mass and $\omega_1$ the
characteristic phonon frequency of wire. The phonon bath and its
bilinear coupling to the wire system is described by
\begin{equation}
\begin{split}
H^{\text{ph}}_{\text{leads}} & +H^{\text{ph}}_{\text{contacts}} \\
&=\sum_{l,k}\left\{\frac{p^2_{lk}}{2m_l}+\frac{m_l\omega^2_{lk}}{2}\left(x_{lk}-\frac{g_{l}Q}{m_l\omega_{lk}}\right)^2\right\}
,
\end{split}
\end{equation}
where $x_{lk},p_{lk},\omega_{lk}$ are the position operators,
momentum operators, and frequencies associated with the bath degrees
of freedom; $m_{l}$ are the masses and $g_l=g_\text{L}=g_\text{R}=g$ represent a
symmetric phonon wire-lead coupling strength for lead $l=\text{L},\text{R}$. The
position and momentum operators can be expressed in terms
of the creation and annihilation operators for phonons as
$x_{lk}=\sqrt{\hbar/2m_l\omega_{lk}}(a^{\dagger}_{lk}+a_{lk})$ and
$p_{lk}=i\sqrt{\hbar m_l\omega_{lk}/2}(a^{\dagger}_{lk}-a_{lk})$.

Throughout the following we assume that slow, time modulated
temperature system acting on the baths are always sufficiently slow
so that a thermal quasi-equilibrium for the molecular wire system
can assumed. The heat transport then obeys the adiabatic, exact
coherent quantum transport laws as discussed in the next subsection,
see Eq. (\ref{j}) and (\ref{jj}) below.

\subsection{Adiabatic modulation of the lead temperatures}

At thermal equilibrium with temperature $T = T_\mathrm{\text{L}} =
T_\mathrm{\text{R}}$ with equal electro-chemical potentials
$\mu_\mathrm{\text{L}} = \mu_\mathrm{\text{R}} = \mu$, the density
matrix for the leads read $\rho_{l}\propto
e^{-(H^{\text{ph}}_{l}+H^{\text{el}}_{l}-\mu_{l}N_{l})/k_\text{B}T_l}$,
where $N_l=\sum_qc^{\dagger}_{lq}c_{lq}$ is the number of electrons
in lead $l = \text{L},\text{R}$ and $k_\text{B}T_l$ denotes the
present lead temperature multipled by the the Boltzmann constant. To
induce shuttling of heat, we invoke a non-equilibrium situation via
an adiabatically slow temperature modulation $T_l(t)$ in the leads.
The latter can be realized experimentally, for example, by use of a
heating/cooling circulator \cite{laser}. Then the expectation values
of the electron and phonon lead operators then read
\begin{align}
\langle c^{\dagger}_{l'q'}c_{lq}\rangle
& = f_l(\epsilon_q,T_l(t))\delta_{ll'}\delta_{qq'}\,,
\\
\langle a^{\dagger}_{l'k'}a_{lk}\rangle
& = n_l(\omega_{k},T_l(t))\delta_{ll'}\delta_{kk'} ,
\end{align}
where
$f_l(\epsilon,T_l(t))=[\text{exp}((\epsilon-\mu_l)/k_\text{B}T_l(t))+1]^{-1}$
and
$n_l(\omega,T_l(t))=[\text{exp}(\hbar\omega/k_\text{B}T_l(t))-1]^{-1}$
denote the Fermi-Dirac distribution and the Bose-Einstein
distribution, respectively, which both inherit a time-dependence from
the temperature modulation. This implies a time-scale separation which
is justified by the fact that laser heating of a metallic system, the
electrons undergo rather fast thermalization
\cite{fast1,fast2,fast3,fast4}.  The corresponding relaxation times
stem from electron-electron and electron-phonon interaction, and for a
typical metal is in the order of a few fs or ps, respectively
\cite{ts1,ts2}.  Therefore, the changes of the lead temperatures occur
on a time-scale much smaller than the thermal fluctuations itself,
i.e.\ $2\pi/\Omega\gg 1$\,ps.

The time-varying lead temperatures $T_{\text{L}}(t)$ and
$T_{\text{R}}(t)$ are assumed to be time-periodic
$T_{\text{L}(\text{R})}(t)=T_{\text{L}(\text{R})}(t+2\pi/\Omega)$,
where $T_0=\overline{T_\text{L}(t)}=\overline{T_\text{R}(t)}$ denotes
the time-averaged environmental reference temperature.
This implies a vanishing temperature bias
\begin{equation}
\overline{\Delta T(t)}\equiv\overline{T_\text{L}(t)-T_\text{R}(t)}=0 \;.
\end{equation}

In the long-time limit, the time-dependent, asymptotic heat current
$J_{Q}(t)=J^{\text{el}}_{Q}(t)+J^{\text{ph}}_{Q}(t)$ assumes the periodicity $2\pi/\Omega$
of the external driving field
\begin{equation}
J_{Q}(t)=J_{Q}(t+{2\pi}/{\Omega}).\label{jp}
\end{equation}
Henceforth we focus on the stationary heat current $\overline{J_{Q}}$
which follows from the average over a full driving period:
\begin{equation}
 \overline{J_{Q}}=\frac{\Omega}{2\pi}\int_0^{2\pi/\Omega}J_{Q}(t)dt\label{javer}.
\end{equation}

If the lead temperatures are modulated slowly enough (adiabatic
temperature rocking), the dynamical thermal bias $\Delta T(t)$ can
be viewed as a static bias at time $t$ in the adiabatic limit
$\Omega\rightarrow 0$. Thus the asymptotic electron and phonon heat
currents $J^{\text{el}(\text{ph})}_{Q}(t)$ can be expressed by
the  Landauer-type formula for electron heat flux \cite{sivan,zheng}
and for the phonon heat flux \cite{rego,ciraci,segal,jswang}, such
that
\begin{align}
 J^{\text{el}}_{Q}(t)=&\frac{1}{2\pi\hbar}\int^{\infty}_{-\infty}
 d\varepsilon(\varepsilon-\mu)\mathcal{T}^{\text{el}}(\varepsilon)\nonumber\\
&\times\left[f(\varepsilon,T_\text{L}(t))-f(\varepsilon,T_\text{R}(t))\right],\label{j}
\\
 J^{\text{ph}}_{Q}(t)=&\frac{1}{2\pi}\int^{\infty}_0 d\omega\hbar\omega\mathcal{T}^{\text{ph}}(\omega)\notag\\
&\times\left[n(\omega,T_\text{L}(t))-n(\omega,T_\text{R}(t))\right],\label{jj}
\end{align}
where $\mathcal{T}^{\text{el}}(\varepsilon)$ and $\mathcal{T}^{\text{ph}}(\omega)$
denote the temperature independent transmission coefficients for
electrons with energy $\varepsilon$ and phonons with frequency
$\omega$ scattered from left lead to right lead, respectively. Note
that the two opposite heat fluxes are not at equilibrium with each
other and that the heat energy transferred by a single electron
scattering process is $\varepsilon-\mu$ rather than $\varepsilon$
\cite{rey}.
The reason for this is the following. At zero temperature, where the
energy levels below Fermi energy $\mu$ are fully occupied, no heat
current is transferred since no electron can tunnel. At finite
temperatures, the tunneling process is thermally activated. An
electron with energy $\varepsilon$ tunneling from left lead to right
lead will dissipate to the Fermi energy level. Therefore the heat
energy transferred by this electron is $\varepsilon-\mu$.

The electron transmission coefficient
$\mathcal{T}^{\text{el}}(\varepsilon)$ can be expressed by the electron
Green's functions \cite{kohler_pr}
\begin{equation}\label{T}
\mathcal{T}^{\text{el}}(\varepsilon)
=\text{Tr}[G^{\dagger}(\varepsilon)\Gamma^\text{R}G(\varepsilon)\Gamma^\text{L}] ,
\end{equation}
where $\Gamma^{l}=|1\rangle\Gamma_l\langle 1|$, stems from the tunnel
coupling to lead $l=\text{L},\text{R}$. For the present case
of a one-site wire, this operator is simply a $1\times 1$-matrix, so
that the Green's function reads
\begin{equation}\label{G}
G(\varepsilon)=
\frac{|1\rangle\langle 1|}{\varepsilon-(E_1-i\Gamma)} ,
\end{equation}
where $\Gamma = \frac{1}{2}(\Gamma_\text{L} + \Gamma_\text{R})$.


For a molecular wire with a single level such as described by
Eq.~(\ref{H-wire}), the electron transmission assumes the
Breit-Wigner form and obeys \cite{kohler_pr}
\begin{equation}
\mathcal{T}^{\text{el}}(\varepsilon)
=\frac{\Gamma^2}{(\varepsilon-E_1)^2+\Gamma^2} , \label{trans}
\end{equation}

where we have assumed symmetric electron wire-lead coupling such that
$\Gamma=\Gamma_\text{L}=\Gamma_\text{R}$.

The phonon transmission coefficient $\mathcal{T}^{\text{ph}}(\omega)$
is evaluated following Ref.~\cite{segal}. As is shown in Appendix A,
it assumes for one phonon mode as well a Breit-Wigner form, i.e., the
temperature-independent phonon transmission probability reads
\begin{equation}\label{tr-ph}
 \mathcal{T}^{\text{ph}}(\omega)=\frac{4\omega^2\gamma^2(\omega)}
 {(\omega^2-\omega^2_1)^2+4\omega^2\gamma^2(\omega)} \,,
\end{equation}
where $\gamma(\omega)=ae^{-\omega/\omega_\text{D}}$. Here
$\omega_\text{D}$ is the Debye cut-off frequency of phonon reservoirs
in the lead and $a=\pi g^2/4mM\omega^3_\text{D}$ incorporating the
phonon wire-lead coupling $g=g_\text{L}=g_\text{R}$.

\subsection{Experimental parameters and physical time scales}

In our numerical investigation we insert the electron wire-lead
tunnel rate $\Gamma=0.11$\,eV, which has been used also to describe
electron tunneling between a phenyldithiol (PDT) molecule and gold
contact \cite{datta2003}. The phonon frequency $\omega_1=1.4\times
10^{14}\, \text{s}^{-1}$ is typical for a carbon-carbon bond
\cite{carbon}. For the Debye cut-off frequency for phonon reservoirs
we use the value for gold which is $\omega_\text{D}=2.16\times
10^{13}\,\text{s}^{-1}$.  The phonon coupling frequency
$a=1.04\times 10^{15}\,\text{s}^{-1}$ is chosen such that the static
thermal conductance assumes the value $50\,\text{pW}\text{K}^{-1}$
which has been measured in experiments with alkane molecular
junctions \cite{dlott}.

These parameters imply physical time scales which are worth being
discussed.  During the dephasing time, electron-phonon interactions
within the wire destroy the electron's quantum mechanical phase.  If
this time is larger than the dwell time, i.e., the time an electron spends
in the wire, the electron transport is predominantly coherent
\cite{Nitzan}. Following Ref.~\cite{Nitzan}, we estimate the dwell
time by the tunneling traversal time $\tau\sim\hbar
[(E_1-\mu)^2+\Gamma^2 ]^{-1/2}$ which for our parameters is of the
order $\tau\sim 5$\,fs and, thus, much shorter than the typical
electron-phonon relaxation time (dephasing time) which is of the
order of 1\,ps. This implies that the electronic motion is
predominantly coherent, so that the electron-phonon interaction
within the wire can be ignored. The phonon relaxation time within
wire can be estimated as $1/a\sim 1$\,fs. Among the above mentioned
time scales, the maximum time scale is the electron-phonon
relaxation time which is in the order of ps. Thus the regime of
validity of our assumption for adiabatic temperature modulations is
justified when the angular driving frequency is much slower the
electron-phonon relaxation rate within lead, i.e. $\Omega \ll
1$\,THz.
\begin{figure}[t]
\includegraphics[width=8cm]{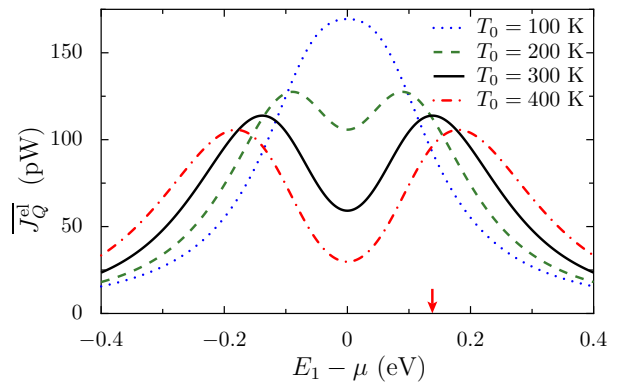}
\caption{(Color online) Directed electronic heat current
$\overline{J^{\text{el}}_{Q}}$ as function of onsite energy
$E_1-\mu$ for different reference temperatures $T_0$ and temperature
oscillation amplitude $A=30$\,K. The arrow marks the onsite energy
$E_1-\mu=0.138$\,eV for which the pumped electron current assumes at
temperature $T_0=300$\,K its maximum.  The adiabatic rocking
frequency is $\Omega = 3.92$\,GHz.} \label{e1}
\end{figure}

\section{Pumping heat via single-sided temperature rocking}

Let us first consider the case in which the temperature of one
lead is modulated sinusoidally, while the temperature of the other
lead is constant,
\begin{align}
\label{tlr}
T_\text{L}(t)&=T_0+A\cos(\Omega t),\\
T_\text{R}(t)&=T_0. \notag
\end{align}
Here $A$ and $\Omega$ are the driving amplitude and (angular)
frequency of the temperature modulation, respectively, while $T_0$ is
the reference temperature. The driving amplitude $A$ is positive and
bounded by the temperature $T_0$ since $T_\text{L}(t)$ has to remain positive
at any time.  The temperature difference between left and right lead
reads $\Delta T(t)=A\cos(\Omega t)$, such that the net thermal bias
vanishes on time-average, $\overline{\Delta T(t)}=0$.

The cycle-averaged heat fluxes, both the electronic and the phononic
one, follow from numerically evaluating the integrals in
Eqs.~(\ref{javer}) and (\ref{j}). As expected for an adiabatic theory,
we observe that the average heat current $\overline{J_{Q}}$ is
independent of the driving frequency $\Omega$. This is in accordance
with the findings for ballistic heat transfer in the adiabatic regime.
\begin{figure}
\centering
 \includegraphics[width=8cm]{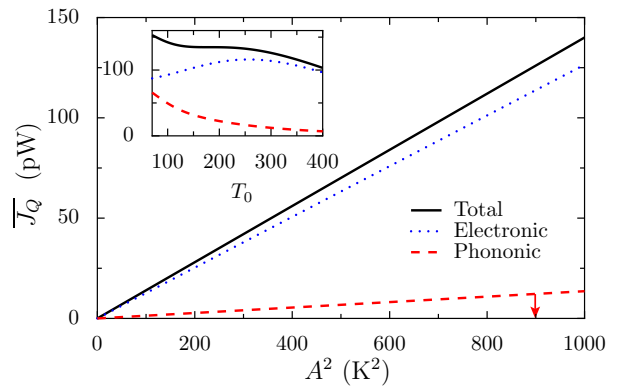}
\caption{(Color online) Total electronic and phononic time-averaged,
directed heat current $\overline{J_{Q}}$ as function of the
squared driving amplitude $A^2$ with reference temperature  at
$T_0=300$\,K for the onsite energy $E_1-\mu=0.138$\,eV. The dotted line
represents the electronic contribution, the dashed line the
phononic one. The inset depicts the
directed heat current as a function of the reference temperature
$T_0$ for the amplitude $A=30$\,K ($A^2=900\,\mathrm{K}^2$) marked by
the arrow in the main panel.} \label{amp}
\end{figure}

In an experiment, the molecular level $E_1$ can be manipulated by
a gate voltage which influences only the electrons.  This allows one
to tune the electron transport while keeping the phonons untouched.
In Fig.~\ref{e1}, we depict the net electron heat current
$\overline{J^{\text{el}}_{Q}}$ as a function of $E_1-\mu$ for
a fixed driving amplitude. We find that the heat current
possesses an extremum for $E_1-\mu = 0$, i.e., when the onsite
energy is aligned with the Fermi energy. Interestingly enough, this
extremum is a maximum for low reference temperature $T_0$ and turns
into a minimum when the temperature exceeds a certain values. This
implies that the net electron heat current is rather sensitive to the
on-site energy with respect to the Fermi energy. This property thus
provides an efficient way to determine experimentally the Fermi energy
of the wire as an alternative to, e.g., measuring the thermopower as
proposed in Ref.~\cite{datta2003}. For large gate variations we find
that the directed electron heat current is significantly suppressed
since the wire level is far off the electron thermal energy, i.e.,
$E_1-\mu\gg k_\text{B}T_0$. The directed heat current then is
dominated by the phonon heat flux. As temperature is increased, the
peak positions of the pumped electron heat current shifts outwards,
away from the Fermi energy. At room temperature $T=300$\,K, the peak
positions are located at $E_1-\mu=\pm\,0.138$\,eV.
\begin{figure}
\centering
 \includegraphics[width=8cm]{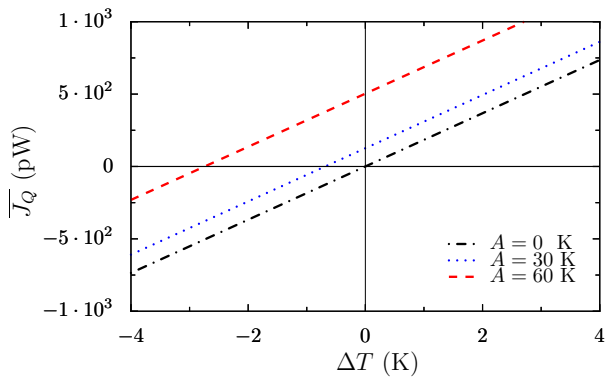}
\caption{(Color online) Total directed heat current $\overline{J_{Q}}$
as the function of static thermal bias $\Delta T$ for  different
driving amplitude strengths $A$ for the temperature modulation. The
reference temperature is set as $T_0=300$ K and the electronic wire
level is set as $E_1-\mu=0.138$ eV.} \label{tbias}
\end{figure}

\subsection{Scaling behavior for small driving strengths}

Figure \ref{amp} shows the total heat current
$\overline{J_{Q}}$ as a function of the driving amplitude $A$ for
the reference temperature $T_0=300$\,K and the electronic site above
the Fermi level. For weak driving ($A\ll T_0$), we find
$\overline{J^{\text{el}(\text{ph})}_{Q}}\propto A^2$ for both
the electronic and the phononic contribution. This behavior can be
understood from a Taylor expansion of the Fermi-Dirac and the
Bose-Einstein distribution,
\begin{equation}
\label{moment}
\begin{split}
g(&\xi,T_\text{L})-g(\xi,T_0)
\\
& = g(\xi,T_0+\Delta T)-g(\xi,T_0)
\\
& =g'(\xi,T_0)\Delta T(t)+\frac{g''(\xi,T_0)}{2}[\Delta T(t)]^2+\cdots
\,,
\end{split}
\end{equation}
where $g$ represents the Fermi-Dirac function $f$ and the
Bose-Einstein function $n$, while $g'$ and $g''$ denote derivatives
with respect to temperature.  Note that the time-dependence stems
solely from the temperature difference $\Delta T(t)$. After a
cycle average over the driving period, the first term in the expansion
vanishes owing to $\overline{\Delta T(t)}=0$. Therefore,
the leading term of the heat current is of second
order, i.e.\ $\propto \overline{[\Delta T(t)]^2}$, which yields
$\overline{J^{\text{el}(\text{ph})}_{Q}}\propto
\int^{2\pi/\Omega}_0 [\Delta T(t)]^2dt\propto A^2$, as observed
numerically. We also plot the directed
heat current as the function of the reference temperature $T_0$ in the
inset of Fig.~\ref{amp}: The directed phonon heat current decreases
monotonically upon increasing the reference  temperature. However, the
emerging total heat current exhibits a relatively flat behavior in a
large temperature range. This is due to the combined effect from
phonons and electrons. At high temperatures, the electron heat flux
dominates the overall directed heat flow.

\subsection{Thermal load characteristics and ratchet-induced thermoelectric voltage}

Thus far we have studied heat pumping in the absence of a static
temperature bias, i.e.\ for $\overline{\Delta T(t)}=0$. We next
introduce a static thermal bias such that a thermal bias $\Delta T:=
\overline{\Delta T(t)}\neq 0$ emerges.  The resulting total directed
heat current $\overline{J_{Q}}$ is depicted in Fig.~\ref{tbias}.  Within
this load curve, we spot a regime with negative static thermal bias
$\Delta T<0$ and positive-valued overall heat flow until $\Delta T$
reaches the stop-bias value, i.e., we find a so-called Brownian
heat-ratchet effect \cite{BHM1,BHM2}. This means that heat can be
directed against a thermal bias from cold to warm like in a
conventional heat pump. The width of this regime scales with the
driving amplitude $A^2$, cf.\ Fig.~\ref{tbias}.
\begin{figure}
\centering
\includegraphics[width=8cm]{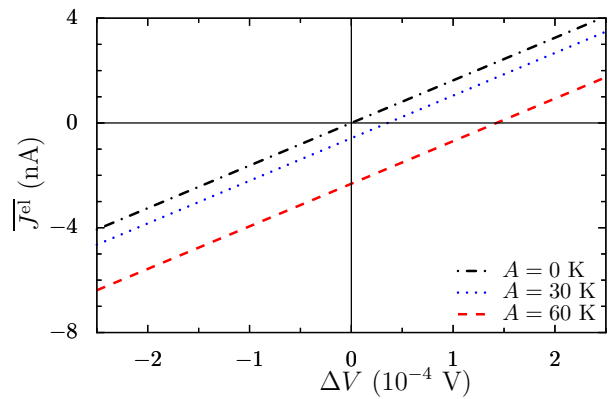}
\caption{(Color online) The time-averaged directed electric current
$\overline{J^{\text{el}}}$ as function of the static voltage bias
$\Delta V$ for different temperature amplitudes $A$. The
reference temperature is $T_0=300$\,K and the electronic wire
level is at $E_1-\mu=0.138$\,eV.} \label{volsee}
\end{figure}

As can be deduced from Fig.~\ref{tbias}, a zero-biased temperature
modulation generates a finite net heat flow at zero temperature bias
similar to the heat flow that would be induced by a static thermal
bias.

Near equilibrium, i.e.\ within the linear response regime, Onsager
symmetry relations for transport of conjugated quantities are
expected to hold. Therefore, the adiabatic temperature modulations
are expected to induce an \textit{electric current} as well. This
net adiabatic electric pump current can be obtained by means of the
cycle-averaged Landauer expression which reads explicitly:
\begin{equation}
\overline{J^{\text{el}}}
=\frac{\Omega}{2\pi}\int_0^{2\pi/\Omega}dt
 \frac{e}{h}\int d\varepsilon\,\mathcal{T}(\varepsilon)
 \left[f(\varepsilon,T_\text{L})-f(\varepsilon,T_\text{R})\right] .
\end{equation}

We in addition apply a net static voltage bias $\Delta V$.
Figure~\ref{volsee}  depicts the net electric current-voltage
characteristics $\overline{J^{\text{el}}}(\Delta V)$ in the presence
of an unbiased temperature modulation while, importantly no external
thermal bias is applied. For a positive bias voltage $\Delta V>0$,
the net electric current is negative, i.e.\ the system effectively
acts as an ``electron pump''.

The value of this externally applied electric stop-voltage $\Delta
V_{\text{st}}$, which renders the electric current vanishing, mimics
here a sole heat-ratchet induced thermopower. In the present
context, this constitutes a novel phenomenon which we term {\it
ratchet Seebeck effect}. Knowingly, the usual thermopower (Seebeck
coefficient) is defined by means of the change in induced voltage
per unit change in applied temperature bias under conditions of zero
electric current \cite{callen}. Here, in the absence of a net thermal
bias, we introduce instead a Gr\"uneisen-like relation, reading
$\gamma=|\Delta V_\text{eff}/\overline{J^\text{el}}|$, where $\Delta
V_\text{eff}$ denotes the effective static voltage bias which yields
the {\it identical} electric heat current $\overline{J^\text{el}}$
as generated by our imposed temperature modulation. It is found that
this effective voltage bias precisely matches the above mentioned
stop-voltage, i.e. $\Delta V_\text{eff}=\Delta V_\text{st}$. Due to
the linear $\overline{J^\text{el}}$--$\Delta V$ characteristics, as
evidenced with Fig.~\ref{volsee}, the Gr\"uneisen-like constant
$\gamma$ is independent of the amplitude of the temperature
modulation. In doing so, we find for $T_0=300$\,K and
$E_1-\mu=0.138$\,eV, that this very Gr\"uneisen-like constant
becomes $\gamma=61.9\times 10^3 \mathrm{V/A}$.
\begin{figure}
\centering
\includegraphics[width=8cm]{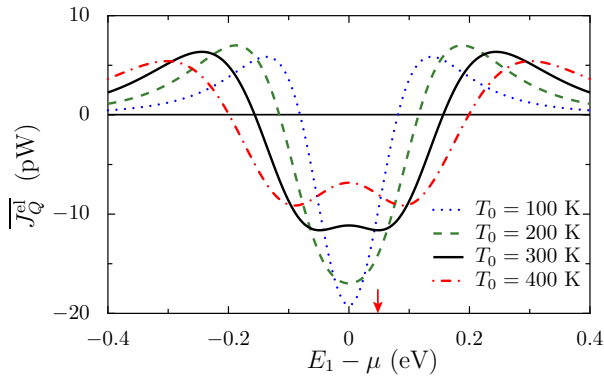}
\caption{(Color online) Directed electron heat current
$\overline{J^{\text{el}}_{Q}}$ as function of the wire level $E_1-\mu$ for
various reference temperatures $T_0$. The driving parameters are
$A_1=A_2=30$\,K and $\varphi=0$.} \label{fige12}
\end{figure}

\section{Temperature rocking in both leads: Pumping heat by
dynamical symmetry breaking}

We next consider temperature modulations applied to both leads in
the absence of a thermal bias. The temperature driving consists of a
contribution with frequency $\Omega$ and a second harmonic with
frequency $2\Omega$. This entails a dynamical symmetry breaking,
namely harmonic mixing \cite{SB1,SB2,SB3,SB4,SB5}. The
time-dependent lead temperatures are chosen as
\begin{align}
T_\text{\text{L},\text{R}} &= T_0 \pm [A_1\cos(\Omega t)-A_2\cos(2\Omega t+\varphi)],
\end{align}
such that again $\overline{T_\text{L}(t)}=\overline{T_\text{R}(t)} = T_0$ and
$\Delta T(t)=2[A_1\cos(\Omega t)+A_2\cos(2\Omega t+\varphi)]$. Then
the average temperature bias vanishes irrespective of the phase
lag $\varphi$.

In Fig.~\ref{fige12}, we depict the resulting electron heat current
$\overline{J^{\text{el}}_{Q}}$ as a function of the on-site
energy $E_1-\mu$ for various reference temperatures $T_0$. At low
temperatures, the net electron heat current exhibits a minimum at
the Fermi energy. With increasing reference temperature $T_0$ this
minimum then develops into a local maximum with two local minima in
its vicinity. The arrow in Fig.~\ref{fige12} marks the minimum at
$E_1-\mu=0.049$\,eV for $T_0=300$\,K. It is interesting that the
direction of the net electron heat current can be tuned by the gate
variation. For an electron wire level close to the Fermi energy, the
directed electron heat current is negative.  By tuning the gate
voltage, the heat current undergoes a reversal and becomes positive
when $E_1-\mu$ is larger than $0.15$\,eV (at reference temperature
$T_0=300$\,K) and eventually approaches zero again for large
detuning.

\begin{figure}
\centerline{\includegraphics[width=8cm]{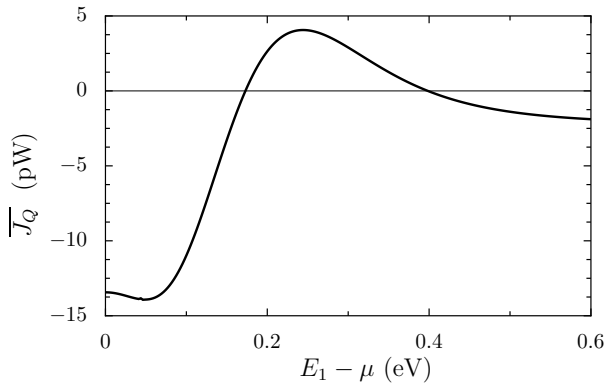}} \caption{Total
net heat current $\overline{J_{Q}}$ as function of the wire level
$E_1-\mu$ for reference temperature $T_0=300$\,K, amplitudes
$A_1=A_2=30$\,K, and phase lag $\varphi=0$.} \label{figep}
\end{figure}

Figure \ref{figep} shows the corresponding sum of electron and
phonon heat flow, i.e., the net heat current $\overline{J_{Q}}$
as a function of wire level $E_1-\mu$. The net phonon heat current
$\overline{J^{\text{ph}}_{Q}}$ is negative for these parameters
(not depicted) and is not sensitive to the gate voltage. As a
consequence, this sum of phonon and electron heat transport,
$\overline{J_{Q}}$, exhibits \textit{multiple current
reversals} as the onsite energy $E_1-\mu$ increases. For small
values of $E_1-\mu$, i.e., close to the Fermi surface, both the
electron and the phonon heat fluxes are negative  and in phase with the
driving. The absolute value of the total heat current assumes its
maximum (at which the heat current is negative). For intermediate
values of $E_1-\mu$, the direction of total net current
$\overline{J_{Q}}$ is reversed due to the dominating positive
contribution of the electrons.  At even larger values of $E_1-\mu$,
the electron heat current almost vanishes, so that the total heat
current is dominated by the negative-valued contribution of the
phonons.  In the limit of large $E_1-\mu$, we find saturation at a
negative value.
\begin{figure}
\centering
 \includegraphics[width=8cm]{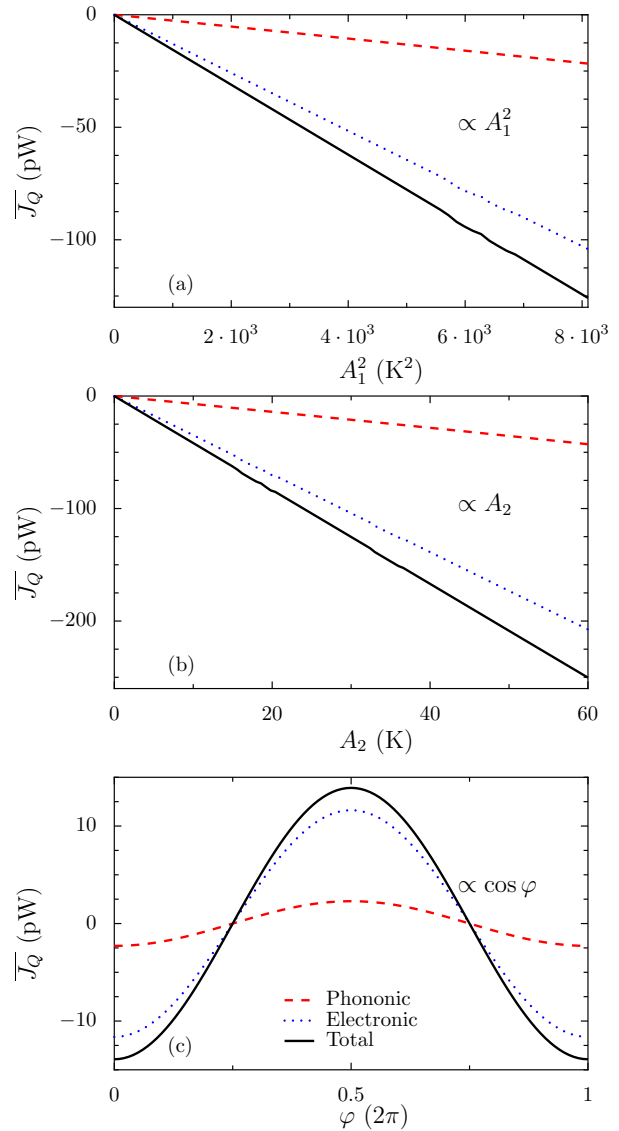}
\caption{(Color online) Heat current $\overline{J_{Q}}$ as function of
(a) fundamental driving strength $A_1$, where $A_2=30$\,K and
$\varphi=0$ and (b) as function of the second-harmonic amplitude
$A_2$, where $A_1=90$\,K and $\varphi=0$. (c) Dependence on the
relative phase $\varphi$ for $A_1=A_2=30$\,K. The reference
temperature is $T_0=300$\,K, while the wire level is
$E_1-\mu=0.049$\,eV.} \label{aphi}
\end{figure}

We also study with  Fig.~\ref{aphi} the net electron and phonon heat
current as the function of driving amplitudes $A_1$, $A_2$ and the
relative phase $\varphi$.  Both contributions scale as
$\overline{J^{\text{el}/\text{ph}}_{Q}}\propto A_1^2A_2\cos(\varphi)$, which implies
that they can be manipulated simultaneously.  This behavior can
be understood by again expanding the Fermi-Dirac and the Bose-Einstein
function at $T_0$:
\begin{equation}
\label{3-moment}
\begin{split}
g(\xi, & T_\text{L})-g(\xi,T_\text{R})
\\
& = g(\xi,T_0+\Delta T/2) - g(\xi,T_0-\Delta T/2)
\\
& = g'(\xi,T_0)\Delta T(t)+\frac{g'''(\xi,T_0)}{24} [\Delta T(t)]^3+\cdots ,
\end{split}
\end{equation}
where $g = f,n$ represents the distribution function for the
electrons and for the phonons, respectively.  Note that the terms of
even order in $\Delta T$ vanish owing to the anti-symmetric
temperature modulation.  Thus, the heat currents are governed by the
time-average of the odd powers $[\Delta T(t)]^{2n+1}$ with $n> 1$,
since $\overline{\Delta T}=0$. It can be easily verified that all
these time-averaged odd moments vanish if either amplitude, $A_1$ or
$A_2$ vanishes.  Note that the lowest lowest-order contribution is
the third moment $\overline{[\Delta T(t)]^3} =
8A_1^2A_2\cos(\varphi)$. Thus, for small driving amplitudes,
$A_1,A_2\ll T_0$, the net electron and phonon heat current are
expected to be proportional to $A_1^2A_2 \cos\varphi$ as is
corroborated with the numerical results depicted in
Figs.~\ref{aphi}(a,b,c). This proportionality to $\cos\varphi$, see
Fig.~\ref{aphi}(c), is even more robust than a priori expected; this
is so  because the cycle averaged 5-th and 7-th moment are
proportional to $\varphi$, as well, i.e., $\overline{[\Delta
T(t)]^5},\overline{[\Delta T(t)]^7}\propto \cos(\varphi)$. This
behavior can be employed for a sensitive control of the heat
current: The direction of the heat current can be reversed by merely
adjusting the relative phase $\varphi$ between the two harmonics.
Note that for the parameters used in the figure, the net electron
heat current $\overline{J^{\text{el}}_{Q}}$ exceeds the net
phonon heat current $\overline{J^{\text{ph}}_{Q}}$ roughly by a
factor $5$.

\section{Conclusions}

We have demonstrated the possibility of steering heat across a gated
two-terminal molecular junction, owing to lead temperatures that
undergo adiabatic, unbiased, time-periodic modulations.
In a realistic molecule, the heat flow is carried by the electrons as
well as by the phonons.  Our study considers both contributions.  Two
scenarios of temperature modulations have been investigated, namely
directed heat flow (i) induced by periodic temperature manipulation
in one connecting lead and (ii) created by a temperature modulation that
includes a contribution oscillating with twice the fundamental
frequency.  In both cases, we predict a finite heat current which is
related to dynamical breaking of reflection symmetry.  A
necessary ingredient is the non-linearity of the initial electron
and phonon distribution, which is manifest in the Fermi-Dirac
distribution and the Bose-Einstein distribution.  The
first scenario yields sizable heat currents proportional to the
squared amplitude of the temperature modulation. The resulting heat
flow occurs in the absence of a static thermal bias. We also studied
heat pumping against an external static thermal bias and computed the
corresponding thermal heat-current load characteristics.
Moreover, the ratchet heat flow in turn generates also an electric
current. This ratchet heat current induces a novel phenomenon,
namely a ratchet-induced, effective thermopower, see in Fig.~\ref{volsee}.

When the asymmetry is induced by temperature rocking at both leads,
the resulting net heat current becomes smaller in size.  This is so
because the leading-order time-averaged heat flow now starts out
with the third moment of the driving amplitude. The benefit of this
second scenario is the possibility of controlling efficiently both
the magnitude and the sign of the net heat flow.  For example, the
direction of the heat current can be readily reversed via the gate
voltage or the relative phase between two temperature modulations that
are harmonically mixed. When adjusting the gate voltage, the directed heat
current experiences multiple current reversals. The directed heat
flow is even up to 7-th order in the amplitude proportional to the
cosine of the phase between the fundamental frequency and the second
harmonic.  This allows robust control even for relatively large
temperature amplitudes.

These theoretical findings may also inspire experimental efforts to
steer heat in a controlled manner across a molecular junction as well
as the development of new concepts for measuring system parameters via
their impact on the heat current. For example, as elucidated in
Sect.~III, the Fermi energy can be sensitively gauged in this way.

\begin{acknowledgments}
The authors like to thank W. H{\"a}usler for his insightful comments
on this work. The work has been supported by the German Excellence
Initiative via the ``Nanosystems Initiative Munich'' (NIM) (P.H.),
the German-Israel-Foundation (GIF) (N.L., P.H.) and the DFG priority
program DFG-1243 ``quantum transport at the molecular scale'' (F.Z.,
P.H., S.K.).
\end{acknowledgments}

\appendix
\section{The derivation of phonon transmission (\ref{tr-ph})}

We derive the phonon transmission coefficient of Eq.~(\ref{tr-ph})
along the lines of Ref.~\cite{segal}. Starting with
Eq.~(33) of that work and assuming symmetric coupling, i.e.\
$\gamma^\text{L}_{k,k'}(\omega_0)=\gamma^\text{R}_{k,k'}(\omega_0)=\gamma_{k,k'}(\omega_0)$,
Eq.~(33) of Ref.~\cite{segal} can be simplified to read
\begin{equation}
\begin{split}
& [\omega^2_k-\omega^2_0+2i\omega_0\gamma_{k,k}(\omega_0)]
  A_k(\omega_0)
\\
&+ i\omega_0\sum_{k'\neq k}\sqrt{\frac{\omega_k}{\omega_{k'}}}
   2\gamma_{k,k'}(\omega_0)A_{k'}(\omega_0)
=\sqrt{\frac{\omega_k}{\omega_0}}V_{0,k}a^{\dagger}_0 ,
\end{split}
\end{equation}
where $\omega_0$ is a dummy variable.

Since we only consider one phonon mode, i.e.\ $k=1$. The second term
in the left hand side of last equation vanishes such that
\begin{equation}\label{ap-a}
 [\omega^2_1-\omega^2_0+2i\omega_0\gamma_{1,1}(\omega_0)]
 A_1(\omega_0)=\sqrt{\frac{\omega_1}{\omega_0}}V_{0,1}a^{\dagger}_0 .
\end{equation}
Substituting Eq.~(46) of Ref.~\cite{segal}, i.e.\
\begin{equation}
 A_k(\omega_0)=\overline{A_k}(\omega_0)V_{0,k}a^{\dagger}_0\sqrt{\frac{\omega_k}{\omega_0}}
\end{equation}
into Eq.~(\ref{ap-a}), we find
\begin{equation}\label{ap-a1}
 \overline{A_1}(\omega_0)=\frac{1}{\omega^2_1-\omega^2_0+2i\omega_0\gamma_{1,1}(\omega_0)}
\end{equation}

For one phonon mode, the phonon transmission is defined from Eq.
(48) in \cite{segal}. However, this definition is $1/2\pi$ times
smaller than the commonly used definition of Ref. \cite{rego} and
\cite{jswang}. With the commonly used definition, the phonon
transmission can be expressed as
\begin{equation}
 \mathcal{T}(\omega)=4\omega^2\gamma^2_{1,1}(\omega)|\overline{A_1}(\omega)|^2
.
\end{equation}
Substituting Eq.~(\ref{ap-a1}) into the last equation and omitting
the subscript in $\gamma_{1,1}$, we obtain
\begin{equation}
 \mathcal{T}(\omega)=\frac{4\omega^2\gamma^2(\omega)}{(\omega^2_1-\omega^2)^2+4\omega^2\gamma^2(\omega)}
\end{equation}
which is the phonon transmission~(\ref{tr-ph}) employed in the main
text.

\end{document}